\documentclass[reprint,prl,aps,showpacs,byrevtex,floatfix]{revtex4-1}
\usepackage{epsf,graphicx,amssymb}
\usepackage[usenames]{color}

\begin{document}

\title{Instability of the origami of a ferrofluid drop in a magnetic field}

\author{Timoth{\'ee} Jamin}
\author{Charlotte Py}
\author{Eric Falcon}
\email[E-mail: ]{eric.falcon@univ-paris-diderot.fr}
\affiliation{Univ Paris Diderot, Sorbonne Paris Cit\'e, MSC, UMR 7057 CNRS, F-75 013 Paris, France}

\date{\today}

\begin{abstract}
Capillary origami is the wrapping of an usual fluid drop by a planar elastic membrane due to the interplay between capillary and elastic forces. Here, we use a drop of magnetic fluid whose shape is known to strongly depend on an applied magnetic field. We study the quasi-static and dynamical behaviors of such a magnetic capillary origami. We report the observation of an overturning instability that the origami undergoes at a critical magnetic field. This instability is triggered by an interplay between magnetic and gravitational energies in agreement with the theory presented here. Additional effects of elasticity and capillarity on this instability are also discussed.
\end{abstract}

%Magnetic fluids and ferrofluids, Static buckling and instability, Liquid-solid interfaces, 
 	
\maketitle
Generally, a solid in contact with a static liquid interface is undeformed by surface tension forces at large scales. However, for sub-millimetric scales or when gravity is negligible, capillary forces may deform an elastic structure in various domains (see \cite{Roman10} for a review): adhesion (coalesence of wet hairs \cite{Bico04}), biological systems (floating flowers \cite{Reis09}), or industrial applications (microscale fabrication such as 3D photovoltaic cells or MEMS \cite{Roman10}). The most well-known phenomenon involving the interplay between elasticity and capillarity is capillary origami. It concerns the spontaneous wrapping of a liquid droplet by a planar elastic sheet when the capillary forces dominate the restoring elastic ones \cite{Py2007,Langre10}. Various folding shapes (spherical, cubic or triangular encapsulation) can then be tuned from the geometry of the initial flat membrane.  A new challenge is to accurately control the folding and unfolding of the origami without the use of mechanical part. Such a capillary origami control has been recently performed by means of an electric field and could lead to potential applications to digital displays \cite{Pineirua10,Yuan10}. However, the capillary origami is generally considered as quasi-static phenomenon and dynamical studies are rare. To our knowledge, only one experiment concerns the dynamical selection of the final shape of the capillary origami when a drop impacts an elastic membrane \cite{Rivetti10}.

In this Letter, a drop of magnetic fluid is deposited on a flat elastic membrane and is submitted to a magnetic field. The shape of the ferrofluid drop is known to strongly depend on the applied magnetic field \cite{Arkhipenko78,Bacri82,Cebers}. The wrapping of the drop by the thin elastic membrane (origami) is thus expected to be strongly modified by the magnetic field. Both quasi-static and dynamical behaviors of such a magnetic capillary origami are reported here.  The most striking one is the observation of an overturning instability that the origami undergoes at a critical magnetic field. This instability is shown to arise from the interplay between magnetic and gravitational energies. 

The ferrofluid used is an ionic aqueous suspension synthesized with 12.4\% by volume of maghemite particles (Fe$_2$0$_3$ ; 7 $\pm 0.3$ nm in diameter) \cite{Talbot}. The properties of this magnetic fluid are: density $\rho=1550$ kg/m$^3$, surface tension $\gamma=43\pm3$ mN/m, initial magnetic susceptibility $\chi_i=0.75$, magnetic saturation $M_{sat}=36\times 10^{3}$ A/m, and dynamic viscosity 1.4 $\times 10^{-3}$ N s/m$^2$. The elastic membranes are made of polydimethylsiloxane (PDMS - Dow Corning Sylgard 184), a 10:1 polymer/curing agent mix. The PDMS is spin coated at rotation rates in the range 1800 -- 2800 rpm over Emery polishing paper (grit 0) with average roughness of 192 $\mu$m. The thickness $h$ of the membrane depends on the rotation rate and ranges from 50 to 100 $\mu$m. Its roughness significantly reduces the adhesion of the membrane on the substrate \cite{Pineirua10}. The ferrofluid drop alone or wrapped by the elastic membrane (origami) is then placed between two horizontal coils, 25  cm (resp. 53 cm) in inner (resp. in outer) diameter, 7 cm far apart. A dc current, $I$, is supplied to the coils in series by a power supply (50 V/50 A). The vertical magnetic induction, $B$, generated is up to 780~G and is measured by a Hall probe located near the drop. The magnitude of $B$ is proportional to $I$ and is controlled either manually or by means of a ramp generator of typical speed rate 0.4 A/s. Deformations of the drop are visualized from a side view by means of a 45 degree mirror reflecting a diffuse lighting towards a high-resolution camera (Pixelink $2208 \times 3000$ pixels) located above the drop, and are recorded up to a 150 Hz sampling.
%The PDMS is then cured in an oven at 65$^{\rm o}$C during 12 hours.  
\begin{figure}[ht!]
    \centering
        \includegraphics[height=65mm]{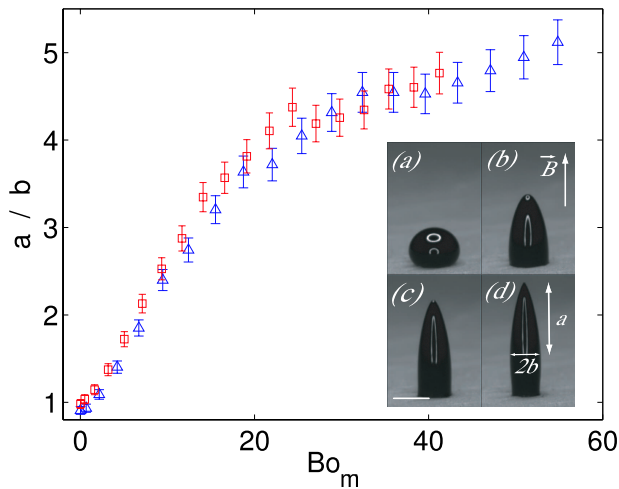}
    \caption{(Color online) Aspect ratio $a/b$ of a ferrofluid drop on a superhydrophobic substrate versus the magnetic Bond number ${\rm Bo_m}$, for two values of the initial drop radius: $R_0=1.2$ ($\bigtriangleup$) and 0.93 mm ($\square$). Inset: Photo of the ferrofluid drop deformation for different vertical magnetic fields: $B=0$ (a), 260 (b), 520 (c), and 780~G (d) corresponding to ${\rm Bo_m}=$ 0, 11, 31, and 53 respectively. $R_0=1.2$ mm. Bar scale is 2 mm.}
    \label{fig01}
\end{figure}

We first study the deformation of a ferrofluid drop alone deposited on a superhydrophobic surface and submitted to a vertical magnetic field. As shown in the inset of Fig.\ \ref{fig01}, the drop is quasi-spherical at $B=0$ (a). When $B$ is increased, the drop lengthens in the direction of $B$ and its shape changes continuously from a semi-ellipsoid (b), then a pointed ellipsoid (c) up to a sharp tip (d). By analogy with a semi-ellipsoid, we define $a$ (resp. $b$) as the semi-major (resp. minor) axis of the deformed drop. $a$ is measured from the top of the drop to the horizontal plane where its width is largest, whereas $b$ is its half-maximal width (see inset of Fig.\ \ref{fig01}). Both are measured on the image of the drop with a 5\% accuracy. The aspect ratio $a / b$ of the drop is plotted in Fig.~\ref{fig01} as a function of the magnetic Bond number, ${\rm Bo_m}$. This dimensionless number characterizes the order of magnitude of the ratio between magnetic and capillary energies \cite{Cebers}. ${\rm Bo_m} \equiv \chi(B)B^2 R_0/ (\mu_0 \gamma$) with $R_0$ the initial drop radius at $B=0$, $\mu_0=4\pi\times 10^{-7}$ H/m the magnetic permeability of the vacuum, and $\chi(B)$ the magnetic susceptibility of the ferrofluid which is a known decreasing function of $B$ with $\chi_i \equiv \chi(B=0)$ \cite{Rosen}. The elongation of the drop is found to strongly increase with ${\rm Bo_m}$ since the ferrofluid tends to align towards the direction of $B$. Moreover, as shown in Fig.\ \ref{fig01}, both curves performed for two different $R_0$ superimpose underlying that a small drop is less deformed than a larger one for the same applied $B$.  Note that such a deformation has been also observed when the ferrofluid drop is surrounded by a non miscible fluid of almost same density leading to a full ellipsoidal shape: $a/b$ is then found to increase continuously with $B$ for small $\chi_i \sim 1$ \cite{Arkhipenko78} whereas for large one ($\chi_i\gtrsim 20$) a discontinuous deformation occurs \cite{Bacri82}. The elongation of a full ellipsoidal droplet can be described theoretically in the limit of a linear \cite{Bacri82,Cebers} or nonlinear \cite{Ivanov11} drop magnetization. To our knowledge, no analogous analytical computation exists for a ferrofluid drop with a semi-ellipsoid shape (as the one in Fig.\ \ref{fig01}). 

\begin{figure}[t!]
 \centering
        \includegraphics[width=85mm]{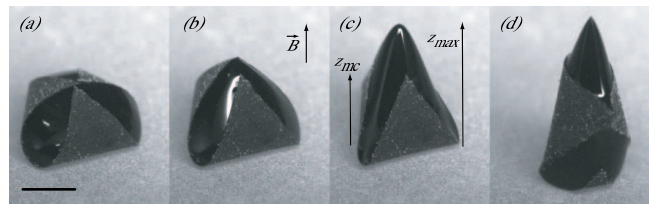}
    \caption{Deformation and instability of the magnetic capillary origami as the vertical magnetic field is increased: $B=0$ (a), 150 (b), 400 (c), and 600~G (d)  corresponding to ${\rm Bo_m}=$ 0, 10, 48, and 84 respectively. An overturning instability occurs between pictures (c) and (d) (see text and movie in Ref.\ \cite{film}). $R_0=2.7$ mm. $L=10$ mm. $h=65~\mu$m. Bar scale is 2~mm.}
    \label{fig02}
\end{figure}

Let us now focus on the behavior of a ferrofluid drop wrapped by an elastic membrane and submitted to a magnetic field. A geometric shape of the membrane is manually cut out from the PDMS layer and placed on the superhydrophobic surface. The membrane shape is chosen to be an equilateral triangle of side $L$ ($5 \leq L \leq 15$ mm) such that the closed state of the origami will have a pyramid-like shape \cite{Py2007}. A typical experiment is as follows. First, at $B=0$, a ferrofluid drop is deposited on the planar membrane. Second, the volume of the drop is adjusted such that, due to the competition between capillary and elastic forces, the three corners of the membrane touch themselves and wrap the spherical drop as shown in Fig.~\ref{fig02}a. When $B$ is turned on and is increased, the drop lengthens vertically (Fig.~\ref{fig02}b), then exhibits a pyramidal shape at higher $B$ due to the presence of the membrane (see Fig.~\ref{fig02}c). For a critical value $B_c$, corresponding to a critical Bond, ${\rm Bo_m^c}$, the origami suddenly overturns, then oscillates around its vertical position before stopping its rocking motion (see movie in \cite{film}). This leads to a new static configuration of the origami (see Fig.~\ref{fig02}d): the drop now forms a cone at its top, and rests on one corner of the membrane, the two other corners wrapping it horizontally. If $B$ is now decreased, the origami falls forward, and never returns to its initial configuration thus showing a hysteretic behavior. 

To quantitatively study the above instability, the maximum height $z_{max}$ of the magnetic capillary origami is measured during a linear temporal increase of $B$ at a tunable rate ranging from 3 G/s to 40 G/s. This rate is slow enough to consider a quasi-static evolution of $B$ (see below). Figure \ref{fig03} shows the dimensionless height $z_{max}/R_0$ as a function of ${\rm Bo_m}$ for the origami and for a ferrofluid drop alone ($R_0$ being the initial radius). For the drop alone,  $z_{max}/R_0$ is a continuous function of ${\rm Bo_m}$. For the origami, a  jump of $z_{max}/R_0$ is observed for a critical Bond number,  ${\rm Bo_m^c}$, corresponding to the threshold of the overturning instability for which the origami shape changes from a pyramid (Fig. \ref{fig02}c) to a cone-like shape (Fig. \ref{fig02}d). Due to inertia, this jump in $z_{max}/R_0$ is accompanied by an overshoot ($z_{max}/R_0$ reaching a maximum higher than the equilibrium) followed by underdamped oscillations (see bottom inset of Fig.\ \ref{fig03} and movie in \cite{film}). Moreover, just after the instability occurring at ${\rm Bo_m^c}$, the origami in its new configuration has a height close to the one of the drop alone meaning that the stress applied by the membrane has relaxed. When iterating this experiment for different sizes $L$ of the triangular membrane of fixed thickness $h$, the onset of this instability ${\rm Bo_m^c}$ is found to increase with $L$, as shown in the top inset of Fig.\ \ref{fig03}. 

\begin{figure}[t!]
 \centering
        \includegraphics[height=60mm]{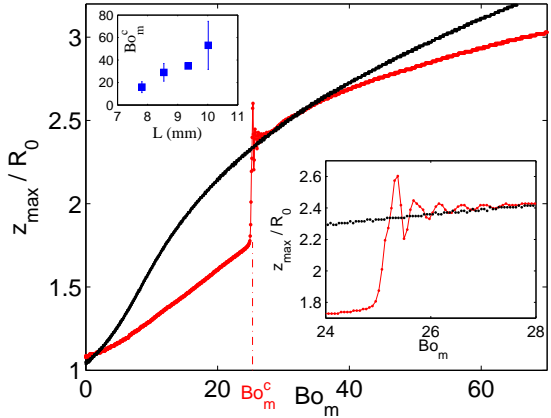}
    \caption{(Color online) Height of the origami [red (light gray) dots] rescaled by its initial radius ($R_0=2.3$ mm), as a function of the magnetic Bond number ${\rm Bo_m}$, and the corresponding evolution for a ferrofluid drop alone (black dots, $R_0=2$ mm). $L=8.4$ mm. $h=56$ $\mu$m. A jump in the height of the origami is observed for a critical Bond ${\rm Bo_m^c}=25.3$, corresponding to the instability shown between Fig. \ref{fig02}c and 2d.  Bottom inset: zoom near the instability region. Top inset: ${\rm Bo_m^c}$ versus the size $L$ of the elastic membrane. }
    \label{fig03}
\end{figure}

We first explain these above observations by dimensional scaling laws. At the onset of the instability, an increase of the origami height is observed (see Fig.\ \ref{fig03}) underlying that gravitational energy $E_g$ should be taken into account. One has $E_g \sim \rho Vgz_{mc}$ with $z_{mc}$ the mass center of the origami and $V$ the volume of the ferrofluid. One can assume $z_{mc} \sim L$, and $V\sim L^3$ since the initial configuration of the origami is quasi-spherical, and thus $E_g\sim L^4$. The magnetic energy writes $E_m \sim \mu_0MHV \sim L^3 B^2\chi(B)/\mu_0$ since the magnetization $M \sim \chi(H)H$ and the magnetic field $H=B/\mu_0$ \cite{Rosen}. The balance between the magnetic and gravitational energies thus leads to the critical magnetic field $B_c$ at the instability threshold such that $\chi(B_c)B_c^2 \sim L$. Thus, ${\rm Bo_m^c} \sim \chi(B_c)B_c^2$ is expected to increase with $L$ as observed in the inset of Fig.\ \ref{fig03}. ${\rm Bo_m^c}$ is also expected to be independent of the membrane thickness $h$ as found experimentally. A similar scaling law analysis balancing the magnetic energy with the capillary energy, $E_c$, or the elastic one, $E_e$, leads for both cases to inverse predictions, i.e. a decrease of ${\rm Bo_m^c}$ with $L$ since $E_c \sim \gamma L^2$, and $E_e \sim Eh^3\kappa^2L^2$  ($E$ being the Young modulus of the membrane of curvature $\kappa \sim 1/L$). Thus, the origami overturning occurring at $B_c$ is triggered by a competition between magnetic and gravitational energies. Note that the role of capillarity and elasticity is non zero. When $B=0$, it drives the origami formation when $E_c \gtrsim E_e$ (i.e. for $L\gtrsim\sqrt{Eh^3/\gamma}$ \cite{Py2007}), and they both deform the ferrofluid shape when $B \neq 0$.

\begin{figure}[t!]
 \centering
        \includegraphics[height=60mm]{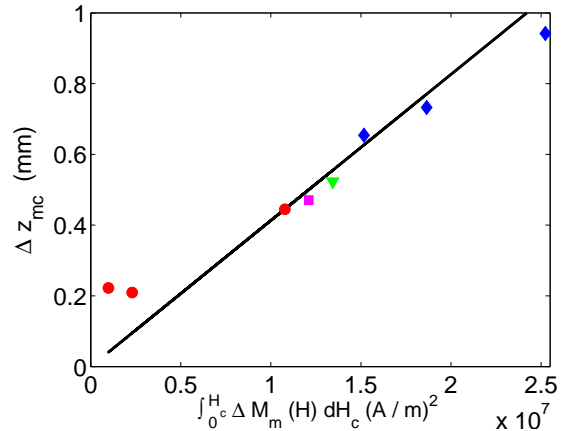}
    \caption{(Color online) Variation of the height of the origami mass center, $\Delta z_{mc}$, during the instability as a function of $\int_0^{H_c}\Delta M_m(H)dH_c$ for different sizes $L$ of the membrane: $L=7.8$ ($\bullet$), 8.5 ($\blacksquare$), 9.4 ($\blacktriangledown$), and 10~mm ($\blacklozenge$). Membrane thickness $h=65$ $\mu$m. Solid line has a slope $\mu_0/(2\rho g)$.}
    \label{fig04}
\end{figure}

Let us now characterize more quantitatively this instability.   To do that, one has to compute the magnetic energy $E_m \equiv -\mu_0V\int_0^{H_0}M(H)dH_0$ with $M$ the ferrofluid magnetization assumed uniform within the drop \cite{Ivanov11}. $H$ is the magnetic field within the ferrofluid droplet that depends on the applied magnetic field $H_0$ through the implicit equation $H=H_0-DM(H)$, $D$ being the demagnetization coefficient that depends on the ferrofluid shape \cite{Rosen}. For a full ellipsoid, it is known that $D=\frac{1-e^2}{2e^3}[\ln{\frac{1+e}{1-e}}-2e]$ with $e=\sqrt{1-(b/a)^2}$ the eccentricity, and $a$ (resp. $b$) the semi-major (resp. minor) axis \cite{Bacri82,Cebers}. For instance, in the case of an ellipsoidal drop alone, $D=0$ for a drop infinitely stretched in the direction of $H_0$, $D=1/3$ for a spherical drop, and $D=1$ for an infinitely flat drop normal to $H_0$. The magnetic energy $E_m$ thus depends on the external magnetic field $H_0$ and on the droplet shape through its magnetization $M(H)$ since $H$ is a function of both $D$ and $H_0$. Consequently, the computation of $E_m$ demands to integrate $M(H)$ over the external magnetic field $H_0$. We estimate $M(H)$ by using the usual Langevin's expression: $M(H) = M_{sat}\mathcal{L}(3\chi_iH/M_{sat})$ where $\mathcal{L}(x)\equiv \coth{(x)}-1/x$ \cite{Rosen}. The initial magnetic susceptibility, $\chi_i\equiv \frac{dM}{dH}|_{H \rightarrow 0}$, and the magnetic saturation, $M_{sat}=M(H)|_{H \rightarrow \infty}$, are given by the ferrofluid properties (see above).

The variation of the magnetic energy during the instability, occurring at the critical field $H_0=H_c$, reads $\Delta E_m=-\mu_0V \int_0^{H_c}\Delta M(H)dH_c$, where $\Delta M \equiv M(H_{2})-M(H_{1})$ is the variation of $M$ before and after the instability (denoted by subscripts $1$ and $2$ respectively). $H_1=H_c-D_1M(H_1)$ and $H_2=H_c-D_2M(H_2)$ are the magnetic field within the drop before and after the instability. Since the origami undergoes a vertical elongation, one has $D_2-D_1<0$ that leads to a decrease of the magnetic energy, i.e. $\Delta E_m < 0$. The variation of the gravitational energy during the instability is $\Delta E_g=\rho Vg\Delta z_{mc}$ with $\Delta z_{mc}$ the variation of the height of the mass center. At the onset of the instability, $\Delta E_m$ compensates $\Delta E_g$, i.e. $\Delta E_m + \Delta E_g=0$, and therefore
\begin{equation}
\Delta z_{mc}=\frac{\mu_0}{\rho g}\int_0^{H_c}\Delta M(H)dH_c {\rm \ .Ê}
\label{theo}
\end{equation}

\begin{figure}[t!]
 \centering
        \includegraphics[height=65mm]{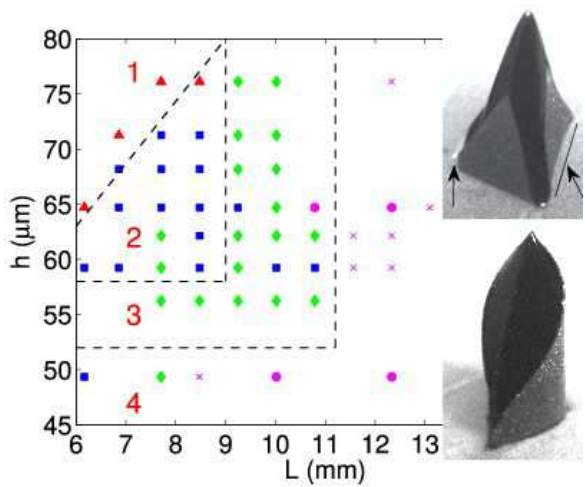}
    \caption{(Color online) Main: Phase diagram of the origami configuration before instability as a function of membrane length, $L$, and thickness, $h$: no origami ($\blacktriangle$), usual origami ($\blacksquare$), origami with dewetting only ($\diamond$), with ridge curvature only ($\bullet$),  with ridge curvature and dewetting ($\times$). Top inset: ferrofluid dewetting (see left arrow) and ridge curvature (see right arrow) before instability. Bottom inset: ``Anomalous'' final state after the instability when adhesion is involved.}
    \label{fig05}
\end{figure}

For different sizes, $L$,  of the membrane, $\Delta z_{mc}$ is deduced from the measurements of $z_{mc}$ before and after the instability occurring at $B_c$. The origami aspect ratio $a/b$ is also measured in the same way as for the ferrofluid drop alone on a plane (see above). For such a semi-ellipsoid, no theoretical expression for the demagnetization coefficient D is known. However, substituting the measured value $a/b$ of the semi-ellipsoid-like origami into the above expression of $D(a/b)$ for a full ellipsoid leads to a measured value of the demagnetization coefficient $D_m$. The quantity $\int_0^{H_c}\Delta M_m(H)dH_c$ is experimentally found as follows ($m$ denoting a measured quantity).  For each $H_c \equiv B_c/\mu_0$ corresponding to each value of $L$, one computes $\Delta M_m(H)$ by using the experimental values of $D_m$ before and after the instability, by numerically solving the above implicit equations given $H_a$ and $H_b$, and finally by iterating $H_0$ from 0 to $H_c$. Fig.\ \ref{fig04} then shows $\Delta z_{mc}$  as a function of $\int_0^{H_c}\Delta M_m(H)dH_c$ for different values of $L$. Both quantities are found proportional, and with a constant $\mu_0/(2\rho g)$, i.e. half the coefficient predicted in Eq.\ (\ref{theo}). Since magnetic energy mostly focuses near drop regions of small curvature (tip-like effect), the value of the magnetic energy for a semi-ellipsoid can be assumed to be half of the one of the full ellipsoid, i.e. $\Delta M=\Delta M_m/2$. The experimental curve is therefore in good agreement with the predictions of Eq.\ (\ref{theo}) with no adjustable parameter. Finally, $\Delta z_{mc}$ is measured to be independent of the thickness $h$ at fixed $L$, emphasizing that elasticity does not play a significant role during this overturning instability. 

%The value of the demagnetization coefficient for a semi-ellipsoid can be assumed to be half of the one of the full ellipsoid, $D=D_m/2$
%If $D=D_m/2$, one has checked that $\Delta M=\Delta M_m/2}

Capillarity and elasticity are at the origin of the capillary origami formation but are not involved in the threshold of  the gravity-magnetic instability reported here. However, capillarity and elasticity effects can appear in the origami configuration before the instability: dewetting of the drop in the vicinity of the strongest curvature of the membrane (see left arrow in the top inset of Fig.\ \ref{fig05}), as well as curvature of a ridge of the pyramid can occur (see right arrow). Both phenomena result from elasticity and capillarity since they depend on the membrane thickness $h$ and length $L$ according to the phase diagram shown in Fig.\ \ref{fig05}. 
For thin enough membranes and/or long enough ones, the elasticity energy $\sim h^3L^0$ is much smaller than the capillary $\sim h^0L^2$ and magnetic $\sim h^0L^3$ ones. It thus leads to very small curvature radii of the membrane that rigidify the origami, and consequently favours the ferrofluid dewetting (see zone 3 in Fig.\ \ \ref{fig05}). For thinner $h$ and/or longer $L$, as elasticity becomes negligible with respect to magnetic effects, the membrane bends inward and follows the elongated ferrofluid shape, leading to the curvature of the pyramid ridges (see zone 4). These additional effects of elasticity and capillarity on the overturning instability deserves further studies. Note also that if an adhesion force exists between the substrate and the membrane, the instability leads to an anomalous final state: one corner of the membrane points up while the two others wrap the drop at its base (see bottom inset of Fig.\ \ref{fig05}).

Finally, the instability reported here could have some interest for potential applications (see Ref. \cite{Roman10}). Indeed, the overturning instability is controlled in a non intrusive way and its scaling law is suitable for miniaturization as lower critical magnetic fields are required for smaller size membranes. 

\begin{acknowledgments}
We thank J.-C. Bacri for fruitful discussions, J. Servais, A. Lantheaume, and M. Pi\~neirua for their technical help. This work has been supported by ANR Turbonde BLAN07-3-197846.
\end{acknowledgments}

%%%%%%%%%%%%%%%%%%%%%%%%%%%%%%%%%%%%%%
%%%%%%%%%%%% REFERENCES %%%%%%%%%%%%%%%%%%
%%%%%%%%%%%%%%%%%%%%%%%%%%%%%%%%%%%%%%

\end{document}